\documentclass[journal=jacsat,manuscript=letter]{achemso}

\usepackage[version=3]{mhchem}
\usepackage{physics}
\usepackage{amsmath}
\usepackage{gensymb}
\usepackage{siunitx}
\usepackage{xcolor}
\author{Sam G. Bishop}
\affiliation{School of Engineering, Cardiff University, Queen's Building, The Parade, Cardiff, UK, CF24 3AA}
\author{John P. Hadden}
\affiliation{School of Physics and Astronomy, Cardiff University, Queen's Building, The Parade, Cardiff, UK, CF24 3AA}
\author{Faris Alzahrani}
\affiliation{School of Physics and Astronomy, Cardiff University, Queen's Building, The Parade, Cardiff, UK, CF24 3AA}
\author{Reza Hekmati}
\affiliation{School of Physics and Astronomy, Cardiff University, Queen's Building, The Parade, Cardiff, UK, CF24 3AA}
\author{Diana L. Huffaker}
\affiliation{School of Engineering, Cardiff University, Queen's Building, The Parade, Cardiff, UK, CF24 3AA}
\alsoaffiliation{School of Physics and Astronomy, Cardiff University, Queen's Building, The Parade, Cardiff, UK, CF24 3AA}
\author{Wolfgang W. Langbein}
\affiliation{School of Physics and Astronomy, Cardiff University, Queen's Building, The Parade, Cardiff, UK, CF24 3AA}
\author{Anthony J. Bennett}
\affiliation{School of Engineering, Cardiff University, Queen's Building, The Parade, Cardiff, UK, CF24 3AA}
\email{BennettA19@cardiff.ac.uk}
\title[Room Temperature Quantum Emitter in Aluminum Nitride]
  {Room Temperature Quantum Emitter in Aluminum Nitride}

\abbreviations{IR,NMR,UV}

\begin{document}




\begin{abstract}
A device that is able to produce single photons is a fundamental building block for a number of quantum technologies. Significant progress has been made in engineering quantum emission in the solid state, for instance using semiconductor quantum dots as well as defect sites in bulk and two dimensional materials. Here we report the discovery of a room temperature quantum emitter embedded deep within the band gap of aluminum nitride (AlN). Using spectral, polarisation and photon-counting time-resolved measurements we demonstrate bright ($>$10\textsuperscript{5} counts $s^{-1}$), pure (g$^2$(0) $<$ 0.2) and polarised room temperature quantum light emission from color centers in this commercially important semiconductor.
\end{abstract}
\section{Keywords}
Aluminum Nitride, Single Photon Source, Quantum Optics, Room Temperature 
\section{Introduction}
Recently, there has been renewed interest in the group III-nitrides as platforms for quantum optics. In the last few years, point-like single photon sources have been reported in hexagonal boron nitride\cite{Tran2016}, gallium nitride (GaN)\cite{Berhane2017,Zhou2018,Nguyen2019}, and very recently in Aluminum Nitride \cite{Xue2020,Lienhard2017}. By virtue of their deep confinement energies, these color centers demonstrate anti-bunching even at room temperature, adding to a select group of solid state materials that host high temperature quantum emitters, such as diamond \cite{Gruber1997,Kurtsiefer2000,Naydenov2011} and silicon carbide (SiC)\cite{Falk2014,Lohrmann2015,Widmann2015}. The commercial applications of the nitrides means there is considerable expertise in processing and the possibility of epitaxial deposition of complex heterostructures, paving the way to cavity enhancement and optoelectronic devices. 

\begin{figure}[!ht]
        \centering
        \includegraphics[width=0.45\linewidth]{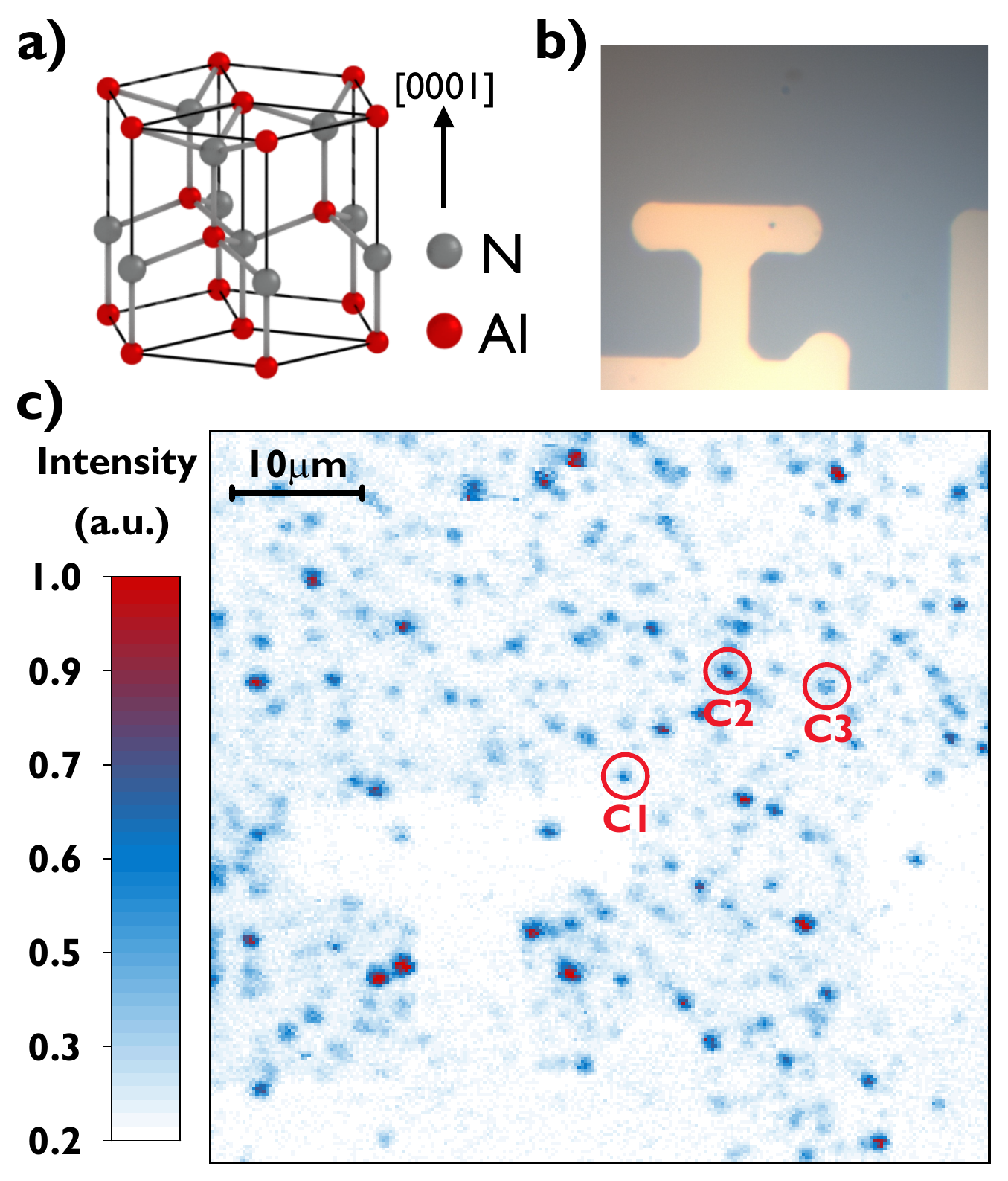}
        \caption{Confocal mapping of point like emitters in AlN. a) Crystal structure of aluminum nitride. The polar crystal axis [0001] is indicated. b) Optical image of the sample with titanium markers. c) Intensity scan map of the sample. The three emitters, C1-3, studied in this work are labelled.}
        \label{fig:Introductory}
\end{figure}

AlN is a semiconductor with a large band gap of \SI{6.2}{\electronvolt}, making it a promising platform for integrated quantum photonic applications \cite{Pernice2012,Xiong2012,Lu2018,Wan2019} due to its transparency in the visible spectrum, strong second-order nonlinearity\cite{Pernice2012}, low-loss high-speed opto-electric phase modulation\cite{Xiong2012}, mature device fabrication and available high-purity substrates. In its wurtzite phase AlN has a hexagonal unit cell, shown in Fig.\ref{fig:Introductory}(a), that lacks inversion-symmetry along the [0001] crystallographic axis. This, along with the finite dipole moment associated with the aluminum-nitrogen bond, leads to internal electric polarisation and piezoelectric effect along [0001]. Recently, AlN has been predicted to host atomic like defects with optically-addressable spin states \cite{Tu2013,Seo2016,Varley2016,Bowes2019}, which would be ideal for quantum technologies that require an interface between flying photonic qubits and stationary trapped spin qubits. The emitters we observe here have a narrower distribution in zero-phonon line energy and a prominent phonon sideband, unlike those in the report of Xue $et$ $al$ \cite{Xue2020}. This suggests the emitters in our sample all arise from the same crystal defect, and have some physical properties consistent with the one emitter reported in the abstract by Lienhard $et$ $al$ \cite{Lienhard2017}.

\section*{Results and Discussion}
 The sample studied within this work is a \SI{1}{\micro\meter} thick epilayer of AlN-on-sapphire. A confocal scanned optical image of the sample shown in Fig.\ref{fig:Introductory}(c) reveals the presence of point-like emitters. A titanium mask on the sample surface is visible, which enables identification of an emitters position. The emitter density is estimated to be \SI{1.25}{\micro\meter^{-2}}, which is sufficient for individual emitters to be optically addressed with a diffraction limited optical spot.

\begin{figure}[!ht]
        \centering
        \includegraphics[width=0.5\linewidth]{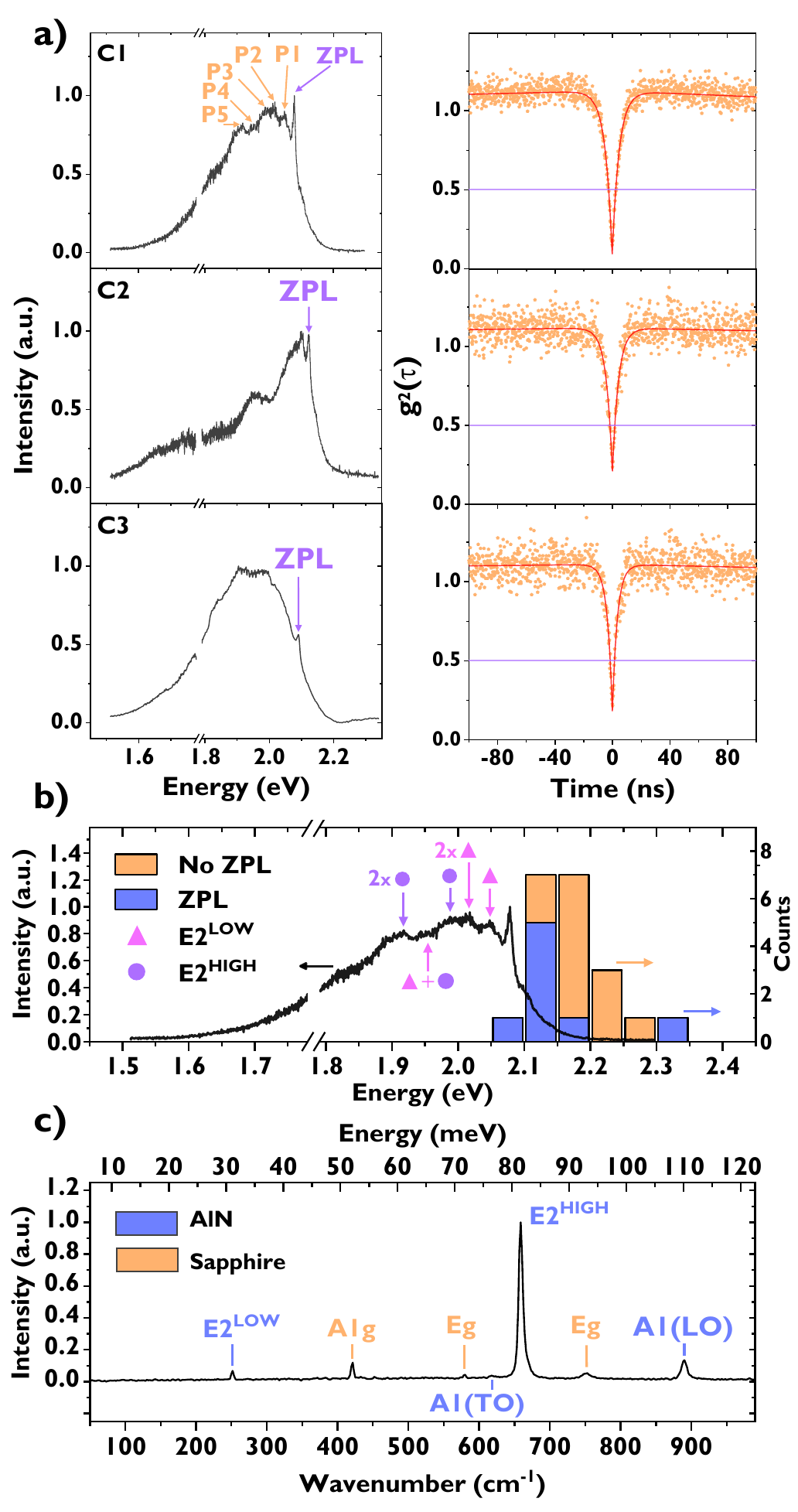}
        \caption{Room temperature spectroscopy of color centers in AlN. a) Spectral and autocorrelation measurements of the three emitters, labelled C1,C2 and C3 in the intensity scan map in Fig.\ref{fig:Introductory}(c). b) Histogram showing the energy in which the spectra for 20 emitters has fallen to half its maximum intensity, on the high energy side. The orange data represents emitters where an obvious zero phonon line cannot be identified. The spectrum for C1 is overlaid, with phonon-shifted energies of the zero phonon line labelled. c) Raman spectrum showing phonon modes in the AlN on sapphire substrate. }
        \label{fig:Statistical}
\end{figure}

 Three color centers, C1-C3, are labelled on the scan map and subsequently studied in detail in Fig.\ref{fig:Statistical}. Spectra of the three emitters shown in Fig.\ref{fig:Statistical}(a) have an obvious spectral peak labelled as a zero phonon line (ZPL) at \SI{2.08}{}, \SI{2.12}{} and \SI{2.09}{\electronvolt}, respectively. This is attributed to the optical excited-to-ground state transition of the color centers without high energy phonon emission, by analogy with the ZPL observed in the NV$^-$ centre in diamond, which has a room temperature ZPL at \SI{1.95}{\electronvolt} \cite{Gruber1997}. In addition to the ZPL peak, each emitters' spectrum demonstrates coupling to high-energy phonons apparent as a broad phonon side band (PSB), spanning about \SI{0.6}{\electronvolt} to the red. Such a broad, prominent, PSB has not been reported in other nitride emitters \cite{Berhane2017,Nguyen2019} but qualitatively resembles that of the NV$^-$ in diamond \cite{Kurtsiefer2000}. 

The room temperature FWHM of the ZPLs are \SI[trapambigerr=false]{8.3\pm0.3}{}, \SI[trapambigerr=false]{11.7\pm0.2}{} and \SI[trapambigerr=false]{9.4\pm0.2}{\milli\electronvolt}, respectively. Broadening of the ZPL at room temperature is consistent with coupling to low energy acoustic phonon modes and spectral jitter. The ZPL contains \SI{3.2}{\%} of the total intensity on average. However, enhancement of the light emitted into the ZPL in a phonon-broadened emitter by weak coupling to an optical cavity has been demonstrated in other materials \cite{Faraon2011}. The spectra for the emitters C2 and C3 have significantly different PSBs. The cause of the change in spectral shapes may result from emitters located in different strain fields or in proximity to crystal dislocations and impurities. 

The quantum nature of the emitters is proven in Fig.\ref{fig:Statistical}(a) which shows the result of Hanbury Brown and Twiss auto-correlation measurements, under \SI{532}{\nano\meter} excitation at \SI{350}{\micro\watt} at room temperature. The emission was filtered between \SI{1.91}{} and \SI{2.33}{\electronvolt}, to suppress the fluorescence signal from the Cr$^3+$ impurity in the sapphire substrate \cite{Jardin1996}. Anti-bunching below the $g^{(2)}(\tau)<0.5$ limit is observed. $g^{(2)}(0)$ for the three emitters C1-3 are \SI[trapambigerr=false]{0.17\pm 0.01}, \SI[trapambigerr=false]{0.20\pm 0.01} and \SI[trapambigerr=false]{0.23\pm 0.02}, respectively.

The phonon energies available within AlN can be determined from a Stokes-shifted Raman measurement, as is shown in Fig.\ref{fig:Statistical}(c). The measurement can provide an insight into Raman-active phonon modes: it is possible to identify a number of vibrational energies from the both AlN and sapphire \cite{Davydov1998,McNeil1993}. The response of the sapphire substrate is to be expected given to the limited axial resolution of the confocal microscope with respect to the AlN thickness. Using the relation $\Delta E=E_{ZPL}-E_{PX}$, where $E_{PX}$ is peak energy for P1-5 in the spectrum for emitter C1, it is possible to correlate the Raman shift due to the vibrational modes from the AlN with the peak locations. Ignoring the contributions due to vibrational modes from the Sapphire, the Raman shifted transitions $\Delta E$; $E2^{(LOW)}$, $A1(TO)$, $E2^{(HIGH)}$ and $A1(LO)$ are given as \SI{31.2\pm0.1}{}, \SI{76.7\pm0.3}{}, \SI{81.7\pm0.01}{} and \SI[trapambigerr=false]{110.3\pm0.1}{\milli\electronvolt} respectively. Therefore, we hypothesise that the peaks P1 and P3 arise due to phonon assisted replicas from coupling to E2$^{(LOW)}$ and E2$^{(HIGH)}$ respectively, as well as peaks P2 and P5 from the corresponding two-phonon processes. In addition, P4 may be described by a mixed two phonon emission involving $E2^{(LOW)}+E2^{(HIGH)}$. These phonon-replica peaks are illustrated in Fig.\ref{fig:Statistical}(b). We hypothesise the higher energy peaks arise from interactions with multiple-phonon-assisted transitions and the broadening from the small defect size, coupling to a large range of the phonon density of states \cite{Davydov1998} in the brillouin zone. Comparable peaks are not as pronounced in the spectra from C2 and C3, due to the contributions merging into a single PSB. The reduced dimensionality of the h-BN system results in a strikingly different phonon-spectrum consisting of replicas of the ZPL spaced by optical photon frequencies with the fraction of the total emission via the ZPL as high as 0.82 \cite{Tran2016}. Conversely previous reports on color centers in GaN\cite{Berhane2017,Nguyen2019} and AlN\cite{Xue2020} do not report high energy phonon sidebands, where the emitters reported here bare a closer resemblance to the NV center in diamond \cite{Kurtsiefer2000}.

A histogram illustrating the energies of 20 emitters, with a bin width of \SI{50}{\milli\electronvolt}, is shown in Fig.\ref{fig:Statistical}(b). All emitters showed a broad emission shape with a ZPL not always being resolved. Therefore, a half max (HM) value is defined which corresponds to the energy at which each spectrum, at the higher energy side, has fallen to half its highest intensity. The same HM value was defined for spectra that have an obvious ZPL (purple) as well as for no obvious ZPL (orange). The histogram illustrates that the majority of the emitters have their HM between \SI{2.10}{} and \SI{2.25}{\electronvolt}, corresponding to ZPLs between \SI{2.00}{} and \SI{2.15}{\electronvolt}. This represents a smaller spectral distribution as compared to color centers recently discovered in GaN, where the ZPL energy between emitters within the same sample varies over \SI{0.4}{\electronvolt}\cite{Berhane2017}. The smaller variation in the ZPL energy for these AlN emitters provides obvious advantage for their exploitation in photonic and/or optoelectronic devices coupled to narrowband cavities or antennae. In addition, it suggests a common origin for all the emitters we have observed in AlN, with energy shifts between emitters resulting from differences in strain, local dislocation density, impurities or point defects. 

\begin{figure}[!ht]
\centering
\includegraphics[width=\linewidth]{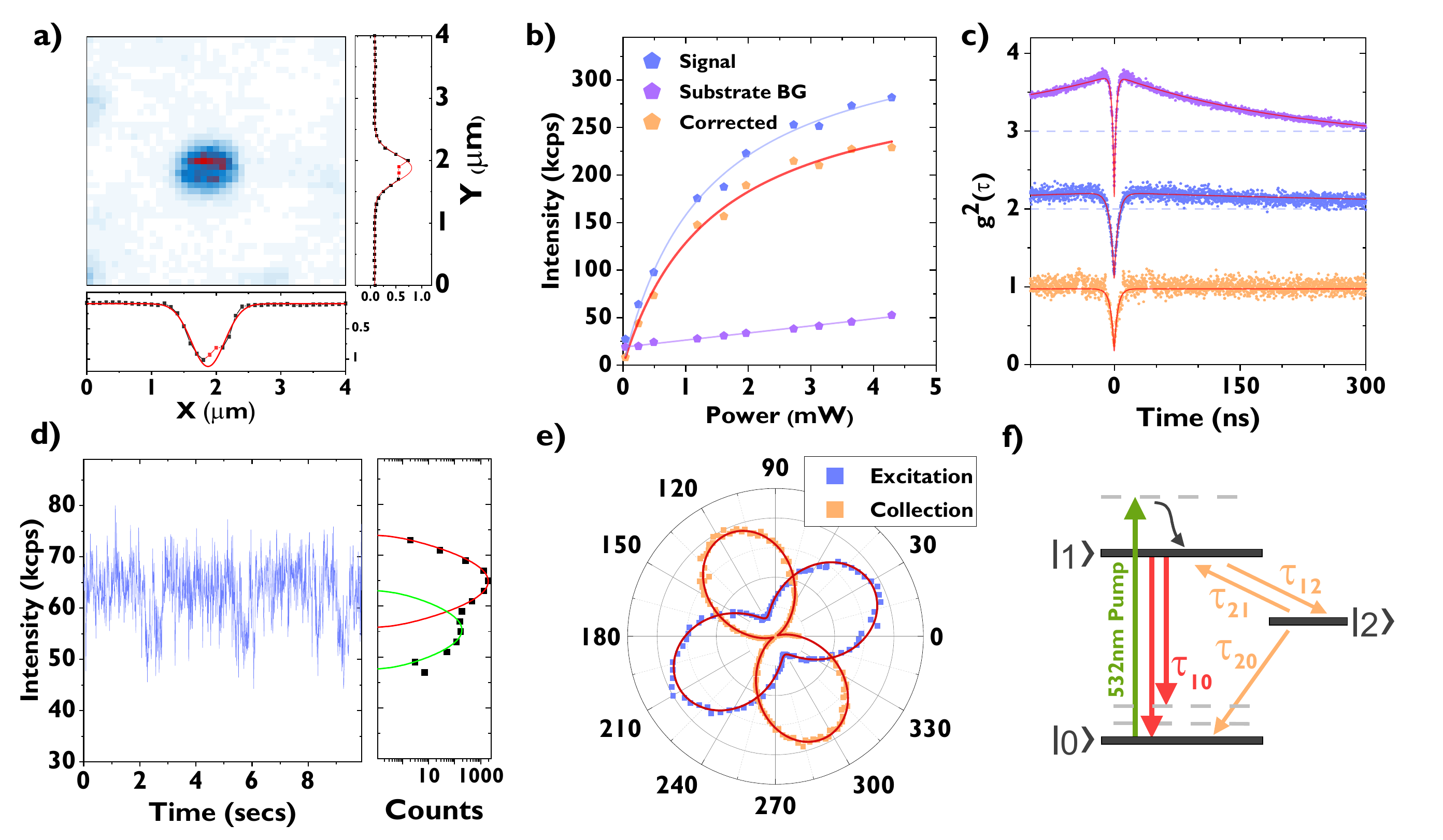}
\caption{Photostability, autocorrelation, polarisation and power dependence of C1 at room temperature. a) Confocal scan over the emitter, with an accompanying X and Y slice. b) Power dependent measurement of intensity showing saturation of intensity at high pump laser power. c) Auto-correlation measurement at a pump power of; 30$\mu$W (orange), 350$\mu$W (blue) and 3.2mW (purple), demonstrating both room temperature anti-bunching and bunching. d) Time resolved stability measurement. e) Polarisation measurement in both excitation and collection. f) Illustration of the three energy level model used to fit the data in c).}
\label{fig:E1Spectroscopy}
\end{figure}

The scan image in Fig.\ref{fig:Introductory}(c) shows that some defects are unstable on a time scale comparable to the 10ms dwell time of each pixel. This is also apparent in the confocal scan around the color centre C1 shown in Fig.\ref{fig:E1Spectroscopy}(a). Intensity slices in both the X and Y axis are presented which demonstrate the diffraction limited size of the emission from the color centre C1. The excitation power (P) dependence of the emission intensity of C1 is shown in Fig.\ref{fig:E1Spectroscopy}(b). The data is fit with the relation $I = I_\infty \times P/(P+P_{sat})$. The saturation intensity $I_{sat}=$\SI{157}{kcps} at $P_{sat} =$ \SI{1.44}{\milli\watt}, which demonstrates the brightness of the emitters at room temperature. The emitter brightness, $>$10\textsuperscript{5} counts $s^{-1}$ at high saturation power, is consistent with other III-Nitride emitters in a high-refractive index bulk \cite{Berhane2017,Xue2020}.

Intensity auto-correlation measurements as a function of excitation power are presented in Fig.\ref{fig:E1Spectroscopy}(c), vertically offset for clarity. Anti-bunching on a \SI[trapambigerr=false]{4.1\pm 0.2}{\nano\second} timescale is observed in parallel with bunching on a \SI[trapambigerr=false]{208.5\pm 0.7}{\nano\second} nanosecond timescale at the highest pump power. The amplitude of the bunching is greater at increasing excitation power, indicative of an internal energy structure that is more complex than a 2-level emitter. We use rate equations \cite{Kurtsiefer2000} to obtain a good fit, by including a meta-stable state, which "shelves" the excitation for a finite time greater than the recombination rate of the excited-to-ground state transition. An illustration of the energy level system is shown in Fig.\ref{fig:E1Spectroscopy}(f), the shelving rate being given by $\tau_{12}$. The purity of the emission is given by $g^{(2)}(0)$ = 0.19$\pm$0.01 at P= 30$\mu$W. Photon bunching at room temperature is also observed in diamond NV centers \cite{Kurtsiefer2000}, SiC \cite{Widmann2015}, h-BN\cite{Tran2016} and GaN color centers\cite{Berhane2017,Zhou2018,Nguyen2019,Xue2020}, with antibunching timescales having the same order of magnitude.

A stability measurement, where the number of photons incident on an avalanche photo-diode is recorded every \SI{10}{\milli\second} for a total of 60 seconds, is presented in Fig.\ref{fig:E1Spectroscopy}(d). The plot shows a 10 second snapshot of the intensity detected, showing intermittent variations between two fluorescence intensities, one centred at \SI{63}{kcps} and the other at \SI{55}{kcps}. A statistical analysis of the intensities seen over the whole 60 second measurement is presented, whereby the intensities are binned into \SI{2}{kcps} bins. The data is fit with two Gaussian distributions with the same width, where the $HWHM\times e^{-1/2}$ variance represents the shot noise of $ \sqrt{n}$ associated with detecting $n$ photons per sampling event. For the more frequently occuring fluorescent state, an average of 650 counts per \SI{10}{\milli\second} are detected, giving a noise value of \SI{2.5}{kcps} in the presented measurement. The occurrence of the less intense fluorescence state is \SI{9}{\%}. This instability is observed in a number of other emitters, but not all, and is reported for other III-Nitride color centers We hypothesise that this instability is not caused by shelving the carrier into the metastable state, as the switching has a time scale orders of magnitude greater than the bunching observed in the $g^{(2)}(t)$ measurement in Fig.\ref{fig:E1Spectroscopy}(c), but rather is a caused by an additional noise source such as a nearby impurity periodically charging and discharging.

A measurement of the emission intensity from C1 as the linear polarisation of the excitation or collection beam are rotated in the plane of the sample is presented in Fig.\ref{fig:E1Spectroscopy}(e). For the collection measurement, the excitation polarisation is aligned to its maximum. Both excitation and collection data demonstrate dipole-like emission patterns. The ratio in intensity between the maximum and minimum is 5:1 and 70:1 for both the excitation and collection, in the absorption and emission, respectively. This linear polarisation observed for all emitters suggests their excitonic dipole is always orientated in the plane of the sample. An angular difference between the maxima in absorption and emission polarisations is consistent with other reports on III-Nitride color centers \cite{Berhane2017,Zhou2018,Xue2020}. We hypothesise that this is due to a multi-level energy structure with excitation and emission transitions having orthogonal polarisation selection rules in the plane of the sample. The absorption and emission dipoles for C1 and C3 are orientated parallel to the [$\bar{1}2\bar{1}0$] plane and m-plane [$10\bar{1}0$] respectively. In addition, polarisation measurements for C2 demonstrate linearly-polarised collection \SI{35}{\degree} offset to the m-plane, along the [$11\bar{2}0$] plane.

\begin{figure}[!ht]
\centering
\includegraphics[width=\linewidth]{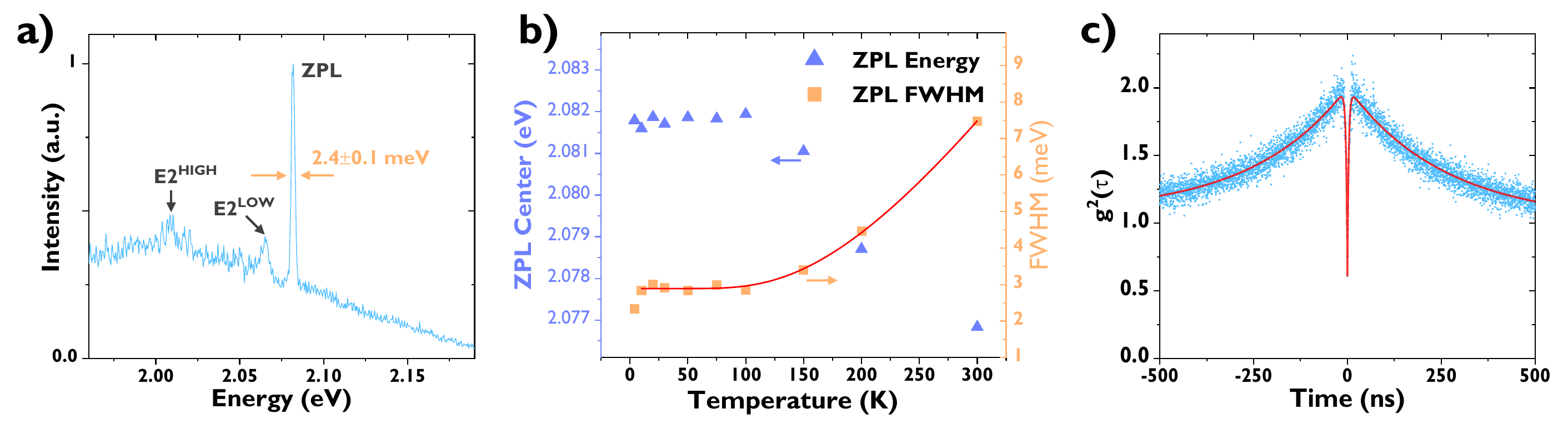}
\caption{Temperature dependence, spectral and autocorrelation measurements of C1 at low temperature. a) Photoluminescence spectrum of C1 at \SI{4}{\kelvin}. b) Temperature dependence of the zero phonon line energy and full width half maximum between \SI{4}{\kelvin} and \SI{300}{\kelvin}. c) Autocorrelation measurement at \SI{4}{\kelvin} under 532 nm excitation at \SI{4}{\milli\watt}.}
\label{fig:LowTemp}
\end{figure}

To gain insight into the temperature dependent properties of the emitters the sample was cooled to \SI{4}{\kelvin} using a closed-cycle helium cryostat. A spectrum of emitter C1 at \SI{4}{\kelvin} is presented in Fig.\ref{fig:LowTemp}(a). A sharpening of the ZPL is observed, as coupling with phonon modes is reduced at lower temperature. The intensity of the emission decreases by a factor of 2 on reducing the temperature from 300 to 4K. 

The ZPL FWHM at \SI{4}{\kelvin} is \SI[trapambigerr=false]{2.4\pm0.1}{\milli\electronvolt}, which is an order of magnitude above the resolution limit of the spectrometer. The temperature dependence of the FWHM (orange) and $E_{ZPL}$ (purple) of the ZPL is presented in Fig.\ref{fig:LowTemp}(b). The FWHM follows a Bose-activated broadening from 10 to \SI{300}{\kelvin}, described by $\gamma = \gamma_0 + \beta/(e^{E/k_B T}-1)$, where $\gamma_0$ is a temperature-independent broadening, $E$ is the activation energy and $\beta$ is the coupling coefficient. A good fit to the data is achieved with the parameters; $\gamma_0$ = \SI[trapambigerr=false]{2.9\pm0.04}{\milli\electronvolt}, $\beta$ = \SI[trapambigerr=false]{29.7\pm0.4}{\milli\electronvolt} and $E$ = \SI[trapambigerr=false]{52.0\pm0.3}{\milli\electronvolt}. The second order correlation measurement in Fig.\ref{fig:LowTemp}(c), taken at \SI{4.0}{\milli\watt} excitation power, confirms the photon statistics at low temperature are comparable to the behaviour measured under ambient conditions (Fig.\ref{fig:E1Spectroscopy}(c)).
\section{Conclusion}

Quantum emission from the sample is attributed to point-like defects embedded deep within the band gap of AlN. The below band gap excitation used here suggests that the defect states are directly being excited. The nature of the emitters is unknown, but several theoretical studies have predicted and studied spin-dependent defects in AlN \cite{Tu2013,Seo2016,Varley2016,Bowes2019}. Secondary ion mass spectrometry (SIMS) measurements from the manufacturer (Dowa Electronics Materials Co.) show trace levels of hydrogen, carbon, oxygen and silicon in the sample. In the future, controlled introduction of other impurities via direct growth or implantation may enable us to engineer desirable spin systems in AlN. Control of the spin states may be possible via the piezoelectric effect, similar to what has been achieved in emitters in SiC \cite{Falk2014}, or through resonant optical fields. The cross-polarised maxima in emission and absorption dipoles presents an ideal arrangement for efficient polarisation filtered resonant control, which conventionally limits the efficiency to \SI{50}{\%} due to the perpendicular excitation and detection optics required to isolate the laser \cite{Wang2019}. Owing to significant existing investment in AlN transducers and sensors this material may be able to compete with diamond and SiC as a viable platform for quantum technologies if it can be shown to host spin-dependent emission from the color centers: in these other materials optical manipulation and read-out of color centre spin states \cite{Neumann2010,Naydenov2011} has enabled sensitive nanoscale sensing \cite{Maze2008,Kraus2014} and promising room temperature qubits.
\section*{Methods}

Room temperature measurements were taken on a confocal microscope with a 0.9NA microscope objective. The collection efficiency is calculated to be \SI{5.1} {\%} into the first lens \cite{Plakhotnik1995}, assuming a dipole orientated in the plane and with a refractive index $n_{AlN}=2.15$ at \SI{600}{\nano\meter}. Low temperature and temperature dependent measurements were taken in an AttoDry 1000 closed-cycle helium cryostat with an internal 0.68NA aspheric lens and a custom optical head. The collection efficiency into the first lens is \SI{2.5}{\percent}. Single mode fibre is used to couple both the excitation and collection into/out of both the room temperature and cryostat microscopes.

Room temperature confocal images were obtained by scanning the laser using dual-axis galvometric mirrors within a optical 4F system. Excitation and collection polarisation measurements was measured using a linear polariser followed by a half-wave plate in the excitation path and a thin film polariser in the collection path. 
 
Time resolved measurements were taken using two SPCM-AQRH silicon avalanche photo-diodes, with unbalanced dark count rates of \SI{50}{} and \SI{520}{cps} respectively. Autocorrelation measurements were taken using a 45:55 fiber beamsplitter. The y-axis offset at time zero in the autocorrelation measurements can be accounted for by including unbalanced dark-count rates, $B_{1}$ and $B_{2}$, on the two detectors, shown in Eq.\ref{eq:DarkCounts}. $S_{1}$ and $S_{2}$ are the signal count rates on the detectors, \SI{4700}{} and \SI{3650}{cps} respectively, and $C_N(\tau)$ the normalised coincidence amplitude at large delay times. After correction, for C1 under 30$\mu$W excitation, we find $g^{(2)}(0) = 0.06$. 
\begin{equation}
    g^{(2)}(\tau) = C_N(\tau)\frac{S_1 S_2 + 2S_1B_2+2S_2B_1+B_1B_2}{S_1S_2} 
    - \frac{2S_1B_2+2S_2B_1+B_1B_2}{S_1S_2}
    \label{eq:DarkCounts}
\end{equation}

A frequency doubled continuous-wave \SI{532}{\nano\meter} Nd:Yag laser was used for excitation in all measurements. Optical filtering was achieved using a ultrasteep \SI{532}{\nano\meter} longpass filter to enable Raman spectroscopy and a short pass 650nm filter to isolate the color centre emission from the Cr$^3+$ emission (\SI{1.78}{\electronvolt}) from the sapphire \cite{Berhane2017}. For spectral measurements beyond 650nm as presented in Fig.\ref{fig:Statistical}(a), the 650nm SP filter is removed and we background correct using a spectrum taken from an area close to the emitter, which represents emission from the substrate. In the future, the study of free-standing AlN or AlN-on-silicon samples would eliminate the need for this spectral filtering. 
\begin{acknowledgement}
The authors thank the financial support provided by the Sêr Cymru National Research Network in Advanced Engineering and Materials, the European Union's Horizon 2020 research and innovation program and the Royal Society for Research Grant RGS\textbackslash R1\textbackslash 191251 and EPSRC grant EP/T017813/1.
\end{acknowledgement}

Since submission of this manuscript, the authors were made aware of a report of single photon emission from defects in AlN by Xue $et$ $al$ \cite{Xue2020}.

\bibliography{library}

\providecommand{\latin}[1]{#1}
\makeatletter
\providecommand{\doi}
  {\begingroup\let\do\@makeother\dospecials
  \catcode`\{=1 \catcode`\}=2 \doi@aux}
\providecommand{\doi@aux}[1]{\endgroup\texttt{#1}}
\makeatother
\providecommand*\mcitethebibliography{\thebibliography}
\csname @ifundefined\endcsname{endmcitethebibliography}
  {\let\endmcitethebibliography\endthebibliography}{}
\begin{mcitethebibliography}{30}
\providecommand*\natexlab[1]{#1}
\providecommand*\mciteSetBstSublistMode[1]{}
\providecommand*\mciteSetBstMaxWidthForm[2]{}
\providecommand*\mciteBstWouldAddEndPuncttrue
  {\def\EndOfBibitem{\unskip.}}
\providecommand*\mciteBstWouldAddEndPunctfalse
  {\let\EndOfBibitem\relax}
\providecommand*\mciteSetBstMidEndSepPunct[3]{}
\providecommand*\mciteSetBstSublistLabelBeginEnd[3]{}
\providecommand*\EndOfBibitem{}
\mciteSetBstSublistMode{f}
\mciteSetBstMaxWidthForm{subitem}{(\alph{mcitesubitemcount})}
\mciteSetBstSublistLabelBeginEnd
  {\mcitemaxwidthsubitemform\space}
  {\relax}
  {\relax}

\bibitem[Tran \latin{et~al.}(2016)Tran, Bray, Ford, Toth, and
  Aharonovich]{Tran2016}
Tran,~T.~T.; Bray,~K.; Ford,~M.~J.; Toth,~M.; Aharonovich,~I. {Quantum emission
  from hexagonal boron nitride monolayers}. \emph{Nature Nanotechnology}
  \textbf{2016}, \emph{11}, 37--41\relax
\mciteBstWouldAddEndPuncttrue
\mciteSetBstMidEndSepPunct{\mcitedefaultmidpunct}
{\mcitedefaultendpunct}{\mcitedefaultseppunct}\relax
\EndOfBibitem
\bibitem[Berhane \latin{et~al.}(2017)Berhane, Jeong, Bodrog, Fiedler,
  Schr{\"{o}}der, Trivi{\~{n}}o, Palacios, Gali, Toth, Englund, and
  Aharonovich]{Berhane2017}
Berhane,~A.~M.; Jeong,~K.-Y.; Bodrog,~Z.; Fiedler,~S.; Schr{\"{o}}der,~T.;
  Trivi{\~{n}}o,~N.~V.; Palacios,~T.; Gali,~A.; Toth,~M.; Englund,~D.;
  Aharonovich,~I. {Bright Room-Temperature Single-Photon Emission from Defects
  in Gallium Nitride}. \emph{Advanced Materials} \textbf{2017}, \emph{29},
  1605092\relax
\mciteBstWouldAddEndPuncttrue
\mciteSetBstMidEndSepPunct{\mcitedefaultmidpunct}
{\mcitedefaultendpunct}{\mcitedefaultseppunct}\relax
\EndOfBibitem
\bibitem[Zhou \latin{et~al.}(2018)Zhou, Wang, Rasmita, Kim, Berhane, Bodrog,
  Adamo, Gali, Aharonovich, and Gao]{Zhou2018}
Zhou,~Y.; Wang,~Z.; Rasmita,~A.; Kim,~S.; Berhane,~A.; Bodrog,~Z.; Adamo,~G.;
  Gali,~A.; Aharonovich,~I.; Gao,~W.-b. {Room temperature solid-state quantum
  emitters in the telecom range}. \emph{Science Advances} \textbf{2018},
  \emph{4}, eaar3580\relax
\mciteBstWouldAddEndPuncttrue
\mciteSetBstMidEndSepPunct{\mcitedefaultmidpunct}
{\mcitedefaultendpunct}{\mcitedefaultseppunct}\relax
\EndOfBibitem
\bibitem[Nguyen \latin{et~al.}(2019)Nguyen, Zhu, Kianinia, Massabuau,
  Aharonovich, Toth, Oliver, and Bradac]{Nguyen2019}
Nguyen,~M.; Zhu,~T.; Kianinia,~M.; Massabuau,~F.; Aharonovich,~I.; Toth,~M.;
  Oliver,~R.; Bradac,~C. {Effects of microstructure and growth conditions on
  quantum emitters in gallium nitride}. \emph{APL Materials} \textbf{2019},
  \emph{7}, 081106\relax
\mciteBstWouldAddEndPuncttrue
\mciteSetBstMidEndSepPunct{\mcitedefaultmidpunct}
{\mcitedefaultendpunct}{\mcitedefaultseppunct}\relax
\EndOfBibitem
\bibitem[Xue \latin{et~al.}(2020)Xue, Wang, Xie, Yang, Xu, Shen, Shi, Jiang,
  Dou, Yu, and Sun]{Xue2020}
Xue,~Y.; Wang,~H.; Xie,~N.; Yang,~Q.; Xu,~F.; Shen,~B.; Shi,~J.-j.; Jiang,~D.;
  Dou,~X.; Yu,~T.; Sun,~B.-q. {Single-Photon Emission from Point Defects in
  Aluminum Nitride Films}. \emph{The Journal of Physical Chemistry Letters}
  \textbf{2020}, \emph{11}, 2689--2694\relax
\mciteBstWouldAddEndPuncttrue
\mciteSetBstMidEndSepPunct{\mcitedefaultmidpunct}
{\mcitedefaultendpunct}{\mcitedefaultseppunct}\relax
\EndOfBibitem
\bibitem[Lienhard \latin{et~al.}(2017)Lienhard, Lu, Jeong, Moon, Iranmanesh,
  Grosso, and Englund]{Lienhard2017}
Lienhard,~B.; Lu,~T.-J.; Jeong,~K.-Y.; Moon,~H.; Iranmanesh,~A.; Grosso,~G.;
  Englund,~D. {High-purity single photon emitter in aluminum nitride photonic
  integrated circuit}. 2017 Conference on Lasers and Electro-Optics Europe {\&}
  European Quantum Electronics Conference. 2017; pp 1--1\relax
\mciteBstWouldAddEndPuncttrue
\mciteSetBstMidEndSepPunct{\mcitedefaultmidpunct}
{\mcitedefaultendpunct}{\mcitedefaultseppunct}\relax
\EndOfBibitem
\bibitem[Gruber \latin{et~al.}(1997)Gruber, Dr{\"{a}}benstedt, Tietz, Fleury,
  Wrachtrup, and von Borczyskowski]{Gruber1997}
Gruber,~A.; Dr{\"{a}}benstedt,~A.; Tietz,~C.; Fleury,~L.; Wrachtrup,~J.; von
  Borczyskowski,~C. {Scanning Confocal Optical Microscopy and Magnetic
  Resonance on Single Defect Centers}. \emph{Science} \textbf{1997},
  \emph{276}, 2012--2014\relax
\mciteBstWouldAddEndPuncttrue
\mciteSetBstMidEndSepPunct{\mcitedefaultmidpunct}
{\mcitedefaultendpunct}{\mcitedefaultseppunct}\relax
\EndOfBibitem
\bibitem[Kurtsiefer \latin{et~al.}(2000)Kurtsiefer, Mayer, Zarda, and
  Weinfurter]{Kurtsiefer2000}
Kurtsiefer,~C.; Mayer,~S.; Zarda,~P.; Weinfurter,~H. {Stable Solid-State Source
  of Single Photons}. \emph{Physical Review Letters} \textbf{2000}, \emph{85},
  290--293\relax
\mciteBstWouldAddEndPuncttrue
\mciteSetBstMidEndSepPunct{\mcitedefaultmidpunct}
{\mcitedefaultendpunct}{\mcitedefaultseppunct}\relax
\EndOfBibitem
\bibitem[Naydenov \latin{et~al.}(2011)Naydenov, Dolde, Hall, Shin, Fedder,
  Hollenberg, Jelezko, and Wrachtrup]{Naydenov2011}
Naydenov,~B.; Dolde,~F.; Hall,~L.~T.; Shin,~C.; Fedder,~H.; Hollenberg,~L.
  C.~L.; Jelezko,~F.; Wrachtrup,~J. {Dynamical decoupling of a single-electron
  spin at room temperature}. \emph{Physical Review B} \textbf{2011}, \emph{83},
  081201\relax
\mciteBstWouldAddEndPuncttrue
\mciteSetBstMidEndSepPunct{\mcitedefaultmidpunct}
{\mcitedefaultendpunct}{\mcitedefaultseppunct}\relax
\EndOfBibitem
\bibitem[Falk \latin{et~al.}(2014)Falk, Klimov, Buckley, Iv{\'{a}}dy,
  Abrikosov, Calusine, Koehl, Gali, and Awschalom]{Falk2014}
Falk,~A.~L.; Klimov,~P.~V.; Buckley,~B.~B.; Iv{\'{a}}dy,~V.; Abrikosov,~I.~A.;
  Calusine,~G.; Koehl,~W.~F.; Gali,~{\'{A}}.; Awschalom,~D.~D. {Electrically
  and Mechanically Tunable Electron Spins in Silicon Carbide Color Centers}.
  \emph{Physical Review Letters} \textbf{2014}, \emph{112}, 187601\relax
\mciteBstWouldAddEndPuncttrue
\mciteSetBstMidEndSepPunct{\mcitedefaultmidpunct}
{\mcitedefaultendpunct}{\mcitedefaultseppunct}\relax
\EndOfBibitem
\bibitem[Lohrmann \latin{et~al.}(2015)Lohrmann, Iwamoto, Bodrog, Castelletto,
  Ohshima, Karle, Gali, Prawer, McCallum, and Johnson]{Lohrmann2015}
Lohrmann,~A.; Iwamoto,~N.; Bodrog,~Z.; Castelletto,~S.; Ohshima,~T.; Karle,~T.;
  Gali,~A.; Prawer,~S.; McCallum,~J.; Johnson,~B. {Single-photon emitting diode
  in silicon carbide}. \emph{Nature Communications} \textbf{2015}, \emph{6},
  7783\relax
\mciteBstWouldAddEndPuncttrue
\mciteSetBstMidEndSepPunct{\mcitedefaultmidpunct}
{\mcitedefaultendpunct}{\mcitedefaultseppunct}\relax
\EndOfBibitem
\bibitem[Widmann \latin{et~al.}(2015)Widmann, Lee, Rendler, Son, Fedder, Paik,
  Yang, Zhao, Yang, Booker, Denisenko, Jamali, Momenzadeh, Gerhardt, Ohshima,
  Gali, Janz{\'{e}}n, and Wrachtrup]{Widmann2015}
Widmann,~M. \latin{et~al.}  {Coherent control of single spins in silicon
  carbide at room temperature}. \emph{Nature Materials} \textbf{2015},
  \emph{14}, 164--168\relax
\mciteBstWouldAddEndPuncttrue
\mciteSetBstMidEndSepPunct{\mcitedefaultmidpunct}
{\mcitedefaultendpunct}{\mcitedefaultseppunct}\relax
\EndOfBibitem
\bibitem[Pernice \latin{et~al.}(2012)Pernice, Xiong, Schuck, and
  Tang]{Pernice2012}
Pernice,~W. H.~P.; Xiong,~C.; Schuck,~C.; Tang,~H.~X. {Second harmonic
  generation in phase matched aluminum nitride waveguides and micro-ring
  resonators}. \emph{Applied Physics Letters} \textbf{2012}, \emph{100},
  223501\relax
\mciteBstWouldAddEndPuncttrue
\mciteSetBstMidEndSepPunct{\mcitedefaultmidpunct}
{\mcitedefaultendpunct}{\mcitedefaultseppunct}\relax
\EndOfBibitem
\bibitem[Xiong \latin{et~al.}(2012)Xiong, Pernice, Sun, Schuck, Fong, and
  Tang]{Xiong2012}
Xiong,~C.; Pernice,~W. H.~P.; Sun,~X.; Schuck,~C.; Fong,~K.~Y.; Tang,~H.~X.
  {Aluminum nitride as a new material for chip-scale optomechanics and
  nonlinear optics}. \emph{New Journal of Physics} \textbf{2012}, \emph{14},
  095014\relax
\mciteBstWouldAddEndPuncttrue
\mciteSetBstMidEndSepPunct{\mcitedefaultmidpunct}
{\mcitedefaultendpunct}{\mcitedefaultseppunct}\relax
\EndOfBibitem
\bibitem[Lu \latin{et~al.}(2018)Lu, Fanto, Choi, Thomas, Steidle, Mouradian,
  Kong, Zhu, Moon, Berggren, Kim, Soltani, Preble, and Englund]{Lu2018}
Lu,~T.-J.; Fanto,~M.; Choi,~H.; Thomas,~P.; Steidle,~J.; Mouradian,~S.;
  Kong,~W.; Zhu,~D.; Moon,~H.; Berggren,~K.; Kim,~J.; Soltani,~M.; Preble,~S.;
  Englund,~D. {Aluminum nitride integrated photonics platform for the
  ultraviolet to visible spectrum}. \emph{Optics Express} \textbf{2018},
  \emph{26}, 11147\relax
\mciteBstWouldAddEndPuncttrue
\mciteSetBstMidEndSepPunct{\mcitedefaultmidpunct}
{\mcitedefaultendpunct}{\mcitedefaultseppunct}\relax
\EndOfBibitem
\bibitem[Wan \latin{et~al.}(2020)Wan, Lu, Chen, Walsh, Trusheim, {De Santis},
  Bersin, Harris, Mouradian, Christen, Bielejec, and Englund]{Wan2019}
Wan,~N.~H.; Lu,~T.~J.; Chen,~K.~C.; Walsh,~M.~P.; Trusheim,~M.~E.; {De
  Santis},~L.; Bersin,~E.~A.; Harris,~I.~B.; Mouradian,~S.~L.; Christen,~I.~R.;
  Bielejec,~E.~S.; Englund,~D. {Large-scale integration of artificial atoms in
  hybrid photonic circuits}. \emph{Nature} \textbf{2020}, \emph{583},
  226--231\relax
\mciteBstWouldAddEndPuncttrue
\mciteSetBstMidEndSepPunct{\mcitedefaultmidpunct}
{\mcitedefaultendpunct}{\mcitedefaultseppunct}\relax
\EndOfBibitem
\bibitem[Tu \latin{et~al.}(2013)Tu, Tang, Zhao, Chen, Zhu, Chu, and
  Fang]{Tu2013}
Tu,~Y.; Tang,~Z.; Zhao,~X.~G.; Chen,~Y.; Zhu,~Z.~Q.; Chu,~J.~H.; Fang,~J.~C. {A
  paramagnetic neutral V Al O N center in wurtzite AlN for spin qubit
  application}. \emph{Applied Physics Letters} \textbf{2013}, \emph{103},
  072103\relax
\mciteBstWouldAddEndPuncttrue
\mciteSetBstMidEndSepPunct{\mcitedefaultmidpunct}
{\mcitedefaultendpunct}{\mcitedefaultseppunct}\relax
\EndOfBibitem
\bibitem[Seo \latin{et~al.}(2016)Seo, Govoni, and Galli]{Seo2016}
Seo,~H.; Govoni,~M.; Galli,~G. {Design of defect spins in piezoelectric
  aluminum nitride for solid-state hybrid quantum technologies}.
  \emph{Scientific Reports} \textbf{2016}, \emph{6}, 20803\relax
\mciteBstWouldAddEndPuncttrue
\mciteSetBstMidEndSepPunct{\mcitedefaultmidpunct}
{\mcitedefaultendpunct}{\mcitedefaultseppunct}\relax
\EndOfBibitem
\bibitem[Varley \latin{et~al.}(2016)Varley, Janotti, and {Van de
  Walle}]{Varley2016}
Varley,~J.~B.; Janotti,~A.; {Van de Walle},~C.~G. {Defects in AlN as candidates
  for solid-state qubits}. \emph{Physical Review B} \textbf{2016}, \emph{93},
  161201\relax
\mciteBstWouldAddEndPuncttrue
\mciteSetBstMidEndSepPunct{\mcitedefaultmidpunct}
{\mcitedefaultendpunct}{\mcitedefaultseppunct}\relax
\EndOfBibitem
\bibitem[Bowes \latin{et~al.}(2019)Bowes, Wu, Baker, Harris, and
  Irving]{Bowes2019}
Bowes,~P.~C.; Wu,~Y.; Baker,~J.~N.; Harris,~J.~S.; Irving,~D.~L. {Space charge
  control of point defect spin states in AlN}. \emph{Applied Physics Letters}
  \textbf{2019}, \emph{115}, 052101\relax
\mciteBstWouldAddEndPuncttrue
\mciteSetBstMidEndSepPunct{\mcitedefaultmidpunct}
{\mcitedefaultendpunct}{\mcitedefaultseppunct}\relax
\EndOfBibitem
\bibitem[Faraon \latin{et~al.}(2011)Faraon, Barclay, Santori, Fu, and
  Beausoleil]{Faraon2011}
Faraon,~A.; Barclay,~P.~E.; Santori,~C.; Fu,~K.-M.~C.; Beausoleil,~R.~G.
  {Resonant enhancement of the zero-phonon emission from a colour centre in a
  diamond cavity}. \emph{Nature Photonics} \textbf{2011}, \emph{5},
  301--305\relax
\mciteBstWouldAddEndPuncttrue
\mciteSetBstMidEndSepPunct{\mcitedefaultmidpunct}
{\mcitedefaultendpunct}{\mcitedefaultseppunct}\relax
\EndOfBibitem
\bibitem[Jardin \latin{et~al.}(1996)Jardin, Canut, and Ramos]{Jardin1996}
Jardin,~C.; Canut,~B.; Ramos,~S. M.~M. {The luminescence of sapphire subjected
  to the irradiation of energetic hydrogen and helium ions}. \emph{Journal of
  Physics D: Applied Physics} \textbf{1996}, \emph{29}, 2066--2070\relax
\mciteBstWouldAddEndPuncttrue
\mciteSetBstMidEndSepPunct{\mcitedefaultmidpunct}
{\mcitedefaultendpunct}{\mcitedefaultseppunct}\relax
\EndOfBibitem
\bibitem[Davydov \latin{et~al.}(1998)Davydov, Kitaev, Goncharuk, Smirnov,
  Graul, Semchinova, Uffmann, Smirnov, Mirgorodsky, and Evarestov]{Davydov1998}
Davydov,~V.~Y.; Kitaev,~Y.~E.; Goncharuk,~I.~N.; Smirnov,~A.~N.; Graul,~J.;
  Semchinova,~O.; Uffmann,~D.; Smirnov,~M.~B.; Mirgorodsky,~A.~P.;
  Evarestov,~R.~A. {Phonon dispersion and Raman scattering in hexagonal GaN and
  AlN}. \emph{Physical Review B} \textbf{1998}, \emph{58}, 12899--12907\relax
\mciteBstWouldAddEndPuncttrue
\mciteSetBstMidEndSepPunct{\mcitedefaultmidpunct}
{\mcitedefaultendpunct}{\mcitedefaultseppunct}\relax
\EndOfBibitem
\bibitem[McNeil \latin{et~al.}(1993)McNeil, Grimsditch, and French]{McNeil1993}
McNeil,~L.~E.; Grimsditch,~M.; French,~R.~H. {Vibrational Spectroscopy of
  Aluminum Nitride}. \emph{Journal of the American Ceramic Society}
  \textbf{1993}, \emph{76}, 1132--1136\relax
\mciteBstWouldAddEndPuncttrue
\mciteSetBstMidEndSepPunct{\mcitedefaultmidpunct}
{\mcitedefaultendpunct}{\mcitedefaultseppunct}\relax
\EndOfBibitem
\bibitem[Wang \latin{et~al.}(2019)Wang, He, Chung, Hu, Yu, Chen, Ding, Chen,
  Qin, Yang, Liu, Duan, Li, Gerhardt, Winkler, Jurkat, Wang, Gregersen, Huo,
  Dai, Yu, H{\"{o}}fling, Lu, and Pan]{Wang2019}
Wang,~H. \latin{et~al.}  {Towards optimal single-photon sources from polarized
  microcavities}. \emph{Nature Photonics} \textbf{2019}, \emph{13},
  770--775\relax
\mciteBstWouldAddEndPuncttrue
\mciteSetBstMidEndSepPunct{\mcitedefaultmidpunct}
{\mcitedefaultendpunct}{\mcitedefaultseppunct}\relax
\EndOfBibitem
\bibitem[Neumann \latin{et~al.}(2010)Neumann, Beck, Steiner, Rempp, Fedder,
  Hemmer, Wrachtrup, and Jelezko]{Neumann2010}
Neumann,~P.; Beck,~J.; Steiner,~M.; Rempp,~F.; Fedder,~H.; Hemmer,~P.~R.;
  Wrachtrup,~J.; Jelezko,~F. {Single-Shot Readout of a Single Nuclear Spin}.
  \emph{Science} \textbf{2010}, \emph{329}, 542--544\relax
\mciteBstWouldAddEndPuncttrue
\mciteSetBstMidEndSepPunct{\mcitedefaultmidpunct}
{\mcitedefaultendpunct}{\mcitedefaultseppunct}\relax
\EndOfBibitem
\bibitem[Maze \latin{et~al.}(2008)Maze, Stanwix, Hodges, Hong, Taylor,
  Cappellaro, Jiang, Dutt, Togan, Zibrov, Yacoby, Walsworth, and
  Lukin]{Maze2008}
Maze,~J.~R.; Stanwix,~P.~L.; Hodges,~J.~S.; Hong,~S.; Taylor,~J.~M.;
  Cappellaro,~P.; Jiang,~L.; Dutt,~M. V.~G.; Togan,~E.; Zibrov,~A.~S.;
  Yacoby,~A.; Walsworth,~R.~L.; Lukin,~M.~D. {Nanoscale magnetic sensing with
  an individual electronic spin in diamond}. \emph{Nature} \textbf{2008},
  \emph{455}, 644--647\relax
\mciteBstWouldAddEndPuncttrue
\mciteSetBstMidEndSepPunct{\mcitedefaultmidpunct}
{\mcitedefaultendpunct}{\mcitedefaultseppunct}\relax
\EndOfBibitem
\bibitem[Kraus \latin{et~al.}(2015)Kraus, Soltamov, Fuchs, Simin, Sperlich,
  Baranov, Astakhov, and Dyakonov]{Kraus2014}
Kraus,~H.; Soltamov,~V.~A.; Fuchs,~F.; Simin,~D.; Sperlich,~A.; Baranov,~P.~G.;
  Astakhov,~G.~V.; Dyakonov,~V. {Magnetic field and temperature sensing with
  atomic-scale spin defects in silicon carbide}. \emph{Scientific Reports}
  \textbf{2015}, \emph{4}, 5303\relax
\mciteBstWouldAddEndPuncttrue
\mciteSetBstMidEndSepPunct{\mcitedefaultmidpunct}
{\mcitedefaultendpunct}{\mcitedefaultseppunct}\relax
\EndOfBibitem
\bibitem[Plakhotnik \latin{et~al.}(1995)Plakhotnik, Moerner, Palm, and
  Wild]{Plakhotnik1995}
Plakhotnik,~T.; Moerner,~W.; Palm,~V.; Wild,~U.~P. {Single molecule
  spectroscopy: maximum emission rate and saturation intensity}. \emph{Optics
  Communications} \textbf{1995}, \emph{114}, 83--88\relax
\mciteBstWouldAddEndPuncttrue
\mciteSetBstMidEndSepPunct{\mcitedefaultmidpunct}
{\mcitedefaultendpunct}{\mcitedefaultseppunct}\relax
\EndOfBibitem
\end{mcitethebibliography}

\end{document}